\documentclass[12pt]{iopart} 
\usepackage{iopams}  
\usepackage{setstack}
\usepackage{graphicx}
\usepackage[applemac]{inputenc}
\begin{document}
%------------------------------------------------------------------------------------------------------------------------------------------
\title{Zipf’s law, power laws, and maximum entropy}
%------------------------------------------------------------------------------------------------------------------------------------------
\author{Matt Visser}
%------------------------------------------------------------------------------------------------------------------------------------------
\address{School of Mathematics, Statistics, and Operations Research\\
Victoria University of Wellington, PO Box 600, \\
Wellington 6140, New Zealand}
%------------------------------------------------------------------------------------------------------------------------------------------
\ead{matt.visser@msor.vuw.ac.nz}
\begin{abstract}
Zipf’s law, and power laws in general, have attracted and continue to attract considerable attention in a wide variety of disciplines --- from astronomy to demographics to software structure to economics to linguistics to zoology, and even warfare. A recent model of \emph{random group formation} [RGF] attempts a general explanation of such phenomena based on Jaynes’ notion of \emph{maximum entropy} applied to a particular choice of \emph{cost function}. In the present article I argue that the specific cost function used in the RGF model is in fact unnecessarily complicated, and that power laws can be obtained in a much simpler way by applying maximum entropy ideas directly to the Shannon entropy subject only to a single constraint: that the average of the logarithm of the observable quantity is specified.   

arXiv:1212.5567

\vskip 10 pt
\noindent 
21 December 2012; 7 January 2013; 10 January 2013; 19 March 2013; \\ \LaTeX-ed \today.

\vskip 10 pt
\noindent 
New Journal of Physics (in press).

\end{abstract}

% Entropy in information theory, 89.70.Cf
% Information theory, 89.70.-a

% phys.soc-ph
% math-ph
% cond-mat.stat-mech

\vspace{2pc}
\noindent

\maketitle

%------------------------------------------------------------------------------------------------------------------------------------------
\bigskip
\hrule
\bigskip
%------------------------------------------------------------------------------------------------------------------------------------------
\markboth{Zipf’s law, power laws, and maximum entropy}{}
\tableofcontents
\markboth{Zipf’s law, power laws, and maximum entropy}{}
%------------------------------------------------------------------------------------------------------------------------------------------
\bigskip
\hrule
\bigskip
%------------------------------------------------------------------------------------------------------------------------------------------
\clearpage
%------------------------------------------------------------------------------------------------------------------------------------------
\markboth{Zipf’s law, power laws, and maximum entropy}{}
%------------------------------------------------------------------------------------------------------------------------------------------
\def\d{{\mathrm{d}}}
\def\O{{\mathcal{O}}}
%------------------------------------------------------------------------------------------------------------------------------------------
\section{Introduction}\label{S:intro}
%------------------------------------------------------------------------------------------------------------------------------------------

Zipf’s law~\cite{Zipf:1935, Zipf:1949, SMS:2009}, and power laws in general~\cite{Newman:2005, Newman:2009, Java}, have and continue to attract considerable attention in a wide variety of disciplines --- from astronomy to demographics to software structure to economics to zoology, and even to warfare~\cite{quarrels}.   Typically one is dealing with integer-valued observables (numbers of objects, people, cities, words, animals, corpses), with $n\in\{1,2,3,\dots \}$. Sometimes the range of values is allowed to be infinite (at least in principle), sometimes a hard upper bound $N$ is fixed (eg, total population if one is interested in subdividing a fixed population into sub-classes). Particularly interesting probability distributions are probability laws of the form:
\begin{itemize}
\item Zipf’s law:  $p_n \propto 1/n$.
\item Power laws: $p_n \propto 1/n^z$.
\item Hybrid geometric/power laws: $p_n\propto w^n/n^z$. 
\end{itemize}
Specifically, a recent model of \emph{random group formation} [RGF], see reference~\cite{RGF:2011}, attempts a general explanation of such phenomena based on Jaynes’ notion of \emph{maximum entropy}~\cite{Jaynes:1957a, Jaynes:1957b, Jaynes:1968, Jaynes:1988, Jaynes:2003} applied to a particular choice of \emph{cost function}~\cite{RGF:2011}. (For recent related work largely in a demographic context see~\cite{Cristelli:2012, Hernando:2012a, Hernando:2012b, Hernando:2012c, Hernando:2012d}. For related work in a fractal context implemented using an  iterative framework see~\cite{barcelona}.) 

In the present article I shall argue that the specific cost function used in the RGF model is in fact unnecessarily complicated, (in fact RGF most typically leads to a hybrid geometric/power law, not a pure power law), and that power laws can be obtained in a much simpler way by applying maximum entropy ideas directly to the Shannon entropy itself~\cite{Shannon:1948, Shannon:1949} subject only to a single constraint: that the average of the logarithm of the observable quantity is specified.  Similarly, I would argue that (at least as long as the main issue one is interested in is “merely” the minimum requirements for obtaining a power law) the appeal to a fractal framework and the iterative model adopted by~\cite{barcelona} is also unnecessarily complicated.

To place this observation in perspective, I will explore several variations on this theme, modifying both the relevant state space and the number of constraints, and will briefly discuss the relevant special functions of mathematical physics that one encounters (zeta functions, harmonic series, poly-logarithms). I shall also discuss an extremely general Gibbs-like model, and the use of non-Shannon entropies (the R\'enyi~\cite{Renyi} and Tsallis~\cite{Tsallis} entropies and their generalizations.) There is a very definite trade-off between simplicity and generality, and I shall very much focus on keeping the discussion as technically simple as possible, and on identifying the simplest model with minimalist assumptions.

\clearpage
%------------------------------------------------------------------------------------------------------------------------------------------
\section{Power laws in infinite state space}\label{S:infinite}
%------------------------------------------------------------------------------------------------------------------------------------------
Let us define the set of observable quantities to be positive integers $n\in\{1,2,3,\dots \}$, without any \emph{a priori} upper bound. The maximum entropy approach~\cite{Jaynes:1957a, Jaynes:1957b, Jaynes:1968, Jaynes:1988, Jaynes:2003}  seeks to estimate the probabilities $p_n$ by maximizing the Shannon entropy~\cite{Shannon:1948, Shannon:1949}, 
\begin{equation}
S=-\sum_n p_n \ln p_n,
\end{equation}
subject to a (small) number of constraints/cost functions --- representing our limited state of knowledge regarding the underlying process.  For example, the RGF model of reference~\cite{RGF:2011} uses one constraint and one (relatively complicated) cost function~\cite{RGF:2011}, in addition to the ``trivial’’ normalization constraint $\sum_n p_n = 1$.  Instead, let us consider the \emph{single} constraint
\begin{equation}
\langle \ln n \rangle \equiv \sum_{n=1}^\infty p_n \ln n = \chi.
\end{equation}
Let us now maximize the Shannon entropy subject to this constraint. This is best done by introducing a Lagrange multiplier $z$ corresponding to the constraint $\langle \ln n \rangle$, plus a second Lagrange multiplier $\lambda$  corresponding to the ``trivial’’ normalization constraint, and considering the quantity:
\begin{equation}
\hat S = - z \left( \sum_{n=1}^\infty p_n \ln n - \chi\right) - \lambda \left( \sum_{n=1}^\infty p_n  -1\right)-\sum_{n=1}^\infty p_n \ln p_n.
\end{equation}
Of course there is no loss of generality in redefining the Lagrange multiplier $\lambda \to Z$ as follows:
\begin{equation}
\hat S = - z \left( \sum_{n=1}^\infty p_n \ln n - \chi\right) - (\ln Z - 1) \left( \sum_{n=1}^\infty p_n  -1\right)-\sum_{n=1}^\infty p_n \ln p_n.
\end{equation}
Varying with respect to the $p_n$ yields the extremality condition
\begin{equation}
- z \ln n - \ln Z - \ln p_n=0,
\end{equation}
with explicit solution
\begin{equation}
p_n = {n^{-z}\over \zeta(z)};  \qquad Z = \zeta(z); \qquad z > 1.
\end{equation}
Here $\zeta(z)$ is the Riemann zeta function~\cite{pure, dirichlet-series, shilov, Havil}, a relatively common and well-known special function, and the condition $z>1$ is required to make the sum $\sum_{n=1}^\infty n^{-z} = \zeta(z)$ converge. This is enough to tell you that one will never \emph{exactly} reproduce Zipf’s law ($z=1$) in this particular manner, though one can get arbitrarily close. The value of the Lagrange multiplier $z$ (which becomes the exponent in the power law) is determined self-consistently in terms of $\chi$ by demanding:
\begin{equation}
\chi(z)  = \langle \ln n \rangle = {\sum_{n=1}^\infty n^{-z} \ln n\over \zeta(z)} = - {\d \zeta(z)/\d z\over\zeta(z)} = - {\d \ln \zeta(z)\over\d z}.
\end{equation}
Then $z\in(1,\infty)$ while $\chi\in(0,\infty)$.  
For practical calculations it is often best to view the exponent $z$ as the single free parameter and $\chi(z)$ as the derived quantity, but this viewpoint can easily be inverted if desired.  
Near $z=1$ we have the analytic estimate
\begin{equation}
\chi(z) = \langle \ln n \rangle = {1\over z-1} - \gamma + \O(z-1),
\end{equation}
where $\gamma$ denotes Euler’s constant.
At maximum entropy, imposing the extremality condition and summing,  we have: 
\begin{eqnarray}
\hat S (z)= S(z) &=& - \sum_{n=1}^\infty  p_n \ln p_n = \ln\zeta(z) + z \chi(z).
\end{eqnarray}
Near $z=1$ we have the analytic estimate
\begin{equation}
\hat S(z) = S(z) = {1\over z-1} +\ln\left({1\over z-1}\right) + 1 - \gamma + \O(z-1).
\end{equation}

For future reference,  it is useful to observe that~\cite[see p 118]{Havil}
\begin{equation}
\zeta(z) = {1\over z-1} + \sum_{m=1}^\infty \gamma_m {(1-z)^m \over m!},
\end{equation}
where the Stieltjes constants $\gamma_m$ satisfy
\begin{equation}
\sum_{n=1}^N {(\ln n)^m\over n} = { (\ln N)^{m+1}\over m+1} + \gamma_m + o(1); \qquad \gamma_0 = \gamma.
\end{equation}
A better estimate, using Euler--Mclaurin summation, is
\begin{equation}
 \sum_{n=1}^N {(\ln n)^m\over n} = {\{\ln(N+{\textstyle{1\over2}} )\}^{m+1}\over m+1}  + \gamma_m + \O\left({\{\ln (N+{\textstyle{1\over2}} )\}^m\over(N+{\textstyle{1\over2}} )^2}\right).
\end{equation}

The quick lesson to take is this: By applying maximum entropy considerations to the \emph{single} constraint $\langle \ln n \rangle =\chi$ you can get a pure power law with \emph{any} exponent $z>1$. 
Furthermore, note that the quantity
\begin{equation}
\exp \langle \ln n \rangle = \prod_{n=1}^\infty n^{p_n}
\end{equation}
is the \emph{geometric mean} of the integers $\{1,2,3,\dots\}$ with the exponents weighted by the probabilities $p_n$.  So one can just as easily obtain the pure power laws considered above by maximizing the entropy subject to the constraint that this geometric mean takes on a specified value. 

\clearpage
%------------------------------------------------------------------------------------------------------------------------------------------
\section{Power laws in finite state space}\label{S:finite}
%------------------------------------------------------------------------------------------------------------------------------------------

If one desires an exact Zipf law, (exponent $z=1$, so that $p_n \propto 1/n$), then because the harmonic series diverges, $\sum_{n=1}^\infty 1/n = \infty$, it is clear that something in the above formulation needs to change.  Perhaps the easiest thing to do is to introduce an explicit maximum value of $n$, call it $N$, so that we take the set of observables to be positive integers $n\in\{1,2,3,\dots, N \}$. (Physicists would call this an infra-red cutoff, or large-distance cutoff.) The maximum entropy approach now amounts to considering
\begin{equation}
\hat S = - z \left( \sum_{n=1}^N p_n \ln n - \chi\right)  - (\ln Z-1) \left( \sum_{n=1}^N p_n  -1\right)-\sum_{n=1}^N p_n \ln p_n.
\end{equation}
Varying with respect to the $p_n$ and maximizing again yields the same extremality condition
\begin{equation}
- z \ln n - \ln Z - \ln p_n=0,
\end{equation}
but now implying
\begin{equation}
p_n = {n^{-z}\over H_N(z)};  \qquad Z= H_N(z).
\end{equation}
Here $H_N(z)$ is the (reasonably well known) generalized harmonic function~\cite{Havil}
\begin{equation}
H_N(z) = \sum_{n=1}^N {1\over n^z}.
\end{equation}
Compared with the previous case the only real difference lies in the normalization function. 
However, because the sum is now always finite, there is no longer any constraint on the value of the exponent $z$, in fact we can have $z\in(-\infty,\infty)$. The case $z=1$ is Zipf’s law, while $z=0$ is a uniform distribution, and $z<0$ corresponds to an ``inverted hierarchy’’ where large values are more common than small values. The price paid for this extra flexibility is that that the model now has two free parameters, which can be chosen to be $z$ and $N$. One has the self-consistency constraint
\begin{equation}
\chi(z, N) = \langle \ln n \rangle= {\sum_{n=1}^N n^{-z} \ln n\over H_N(z)} = - {\d H_N(z)/\d z\over H_N(z)} = - {\d \ln H_N(z)\over\d z}.
\end{equation}
It is easy to check that $\chi$ is now bounded by $\chi\in(0,\ln N)$.  At maximum entropy we now have:
\begin{eqnarray}
\hat S (z,N)= S(z,N) &=& - \sum_{n=1}^N  p_n \ln p_n  = \ln H_N(z) + z \chi(z,N).
\end{eqnarray}
Because this is now a two-parameter model, it will always (naively) be a ``better’’ fit to observational data than a single-parameter model. Sometimes (for $z\leq 1$) retreating to this 2-parameter model is necessary,  but for $z>1$ the one-parameter model of the previous section ($N\to\infty$) should be preferred.

%------------------------------------------------------------------------------------------------------------------------------------------
\section{Zipf’s law in finite state space}\label{S:zipf}
%------------------------------------------------------------------------------------------------------------------------------------------

If  for observational or theoretical reasons one is \emph{certain} that $z=1$, (Zipf’s law), then the model reduces as follows: The state space is $n\in\{1,2,3,\dots, N \}$ where $N$ is now the \emph{only} free parameter. Then explicitly forcing $z\to1$ one considers
\begin{equation}
\hat S = -  \left( \sum_{n=1}^N p_n \ln n - \chi\right)  - (\ln Z-1) \left( \sum_{n=1}^N p_n  -1\right)-\sum_{n=1}^N p_n \ln p_n.
\end{equation}
This is completely equivalent to considering
\begin{equation}
\hat S =  \chi - (\ln Z-1) \left( \sum_{n=1}^N p_n  -1\right)-\sum_{n=1}^N p_n \ln (n \; p_n),
\end{equation}
but writing the quantity to be maximized in this way hides the role of the Shannon entropy. 
Varying with respect to the $p_n$ and maximizing now yields a (very) slightly different extremality condition
\begin{equation}
- \ln n - \ln Z - \ln p_n=0,
\end{equation}
and so
\begin{equation}
p_n = {1\over H_N} {1\over n}; \qquad Z = H_N.
\end{equation}
Here $H_N$ is the (ordinary) harmonic number~\cite{Havil}
\begin{equation}
H_N = \sum_{n=1}^N {1\over n}.
\end{equation}
Then
\begin{equation}
\chi(N) = \langle \ln n \rangle= {1\over H_N}\; \sum_{n=1}^N { \ln n\over n}.
\end{equation}
Furthermore, at maximum entropy
\begin{eqnarray}
\hat S (N)= S(N) &=& -  \sum_{n=1}^N  p_n \ln p_n=  \ln H_N + \chi(N).
\end{eqnarray}
Now we have already seen
\begin{equation}
H_N = \ln (N +{\textstyle{1\over2}} ) + \gamma +  \O\left({1\over(N+{\textstyle{1\over2}} )^2}\right),
\end{equation}
and
\begin{equation}
 \sum_{n=1}^N {\ln n\over n} = {1\over2}\{\ln(N+{\textstyle{1\over2}} )\}^2 + \gamma_1 + \O\left({\ln (N+{\textstyle{1\over2}} )\over(N+{\textstyle{1\over2}} )^2}\right).
\end{equation}
Therefore
\begin{equation}
\chi(N) = \langle \ln n \rangle= {1\over 2} \ln(N+{\textstyle{1\over2}} )+ o(1),
\end{equation}
and at maximum entropy
\begin{equation}
\hat S(N) = S(N) = {1\over 2} \ln(N+{\textstyle{1\over2}} ) + \ln\ln(N+{\textstyle{1\over2}}) + o(1).
\end{equation}
Note that you can use this to estimate the size $N$ of the state space one needs to adopt in order to be compatible with the (observed) logarithmic average $\langle \ln n \rangle$. Indeed
\begin{equation}
N \approx \exp\{ 2 \langle \ln n \rangle \}  =   \{\exp \langle \ln n \rangle \}^2.
\end{equation}
This relates the size of the required state space $N$ to the \emph{square} of the geometric mean $\exp \langle \ln n \rangle$.

%------------------------------------------------------------------------------------------------------------------------------------------
\section{Hybrid geometric/power models in infinite state space}\label{S:hybrid}
%------------------------------------------------------------------------------------------------------------------------------------------
\def\Li{{\mathrm{Li}}}
To generate a hybrid geometric/power law model, similar in output to the RGF model~\cite{RGF:2011}, but with considerably simpler input assumptions, simply take \emph{both} the logarithmic average $\langle \ln n\rangle=\chi$,
\emph{and} the arithmetic average $\langle n\rangle=\mu$, to be specified --- and then maximize the Shannon entropy subject to these two constraints, (plus the trivial normalization constraint). That is, introduce two Lagrange multipliers $z$ and $w$, and maximize
\begin{eqnarray}
\hat S &=& - z \left( \sum_{n=1}^\infty p_n \ln n - \chi\right) + \ln w \left( \sum_{n=1}^\infty p_n n - \mu\right)  
\nonumber\\
&& \qquad - (\ln Z-1)  \left( \sum_{n=1}^\infty p_n  -1\right)
-\sum_{n=1}^\infty p_n \ln p_n.
\end{eqnarray}
Varying with respect to the $p_n$ yields
\begin{equation}
- z \ln n + n \ln w - \ln Z - \ln p_n = 0,
\end{equation}
with solution
\begin{equation}
p_n = {w^n \; n^{-z}\over \Li_z(w)}; \qquad Z = \Li_z(w);  \qquad w < 1.
\end{equation}
Here the normalizing constant is the well-known poly-logarithm function~\cite{pure, dirichlet-series, shilov, Havil}
\begin{equation}
\Li_z(w) = \sum_{n=1}^\infty {w^n\over n^z};  \qquad   \Li_1(w) = -\ln(1-w).
\end{equation}
Then
\begin{equation}
\chi(w,z) \equiv \langle \ln n \rangle= {1\over \Li_z(w)}\; \sum_{n=1}^\infty { w^n \ln n\over n^z} = - {\d \ln \Li_z(w)\over \d z},
\end{equation}
while
\begin{equation}
\mu(w,z) \equiv \langle n \rangle= {1\over \Li_z(w)}\; \sum_{n=1}^\infty { w^n n\over n^z} =  {\Li_{z-1}(w)\over \Li_z(w)}   = {\d \ln \Li_z(w)\over \d \ln w}.
\end{equation}
Furthermore, at maximum entropy
\begin{equation}
\fl
\hat S (w,z)= S(w,z) = -\sum_{n=1}^\infty  p_n \ln p_n 
=  \ln \Li_z(w) + z\; \chi(w,z) -  \mu(w,z)\; \ln w .
\end{equation}
Note that the probability function arising in this model is fully as general as that arising in the RGF model~\cite{RGF:2011}, but with what is perhaps a somewhat clearer interpretation. 

This is because the RGF model uses what may be viewed as an unnecessarily complicated ``cost function’’, with an unnecessary degeneracy in the parameters.  Indeed, from reference~\cite{RGF:2011} one sees (in their notation, slightly different from current notation)
\begin{equation}
I_\mathrm{cost} = \sum_k  P(k) \ln[ k N(k)]; \qquad  P(k) = N(k)/N.
\end{equation}
That is
\begin{eqnarray}
I_\mathrm{cost} &=& \sum_k  P(k) \ln[ k N P(k)] 
\\
&=& \sum_k P(k) \ln k + \sum_k P(k) \ln N + \sum_k P(k) \ln P(k) 
\\
&=&\langle \ln k\rangle + \ln N - S.
\end{eqnarray}
So the RGF cost function~\cite{RGF:2011} is simply a linear combination of Shannon entropy, the logarithmic mean $\langle \ln k\rangle$, and a redundant constant offset $\ln N$. (Unfortunately the $N$ of reference~\cite{RGF:2011} is not the same as the $N$ used in this article, the RGF parameter $N$ corresponds to the number of independent realizations of the underlying statistical process one considers --- it is the number of simulations, or number of universes in the statistical ensemble). 
The additional complexity implicit in the RGF model can to some extent be viewed as a side-effect of forcing data into discrete ordinal boxes when one does not necessarily have good physical/mathematical/demographic reasons for knowing which particular boxes are “best”, how the boxes are to be assigned ordinal numbers, and where the box boundaries should most profitably be placed. 
Apart from the issues raised above, one could in addition explicitly restrict the state space to be finite, adding yet another free parameter, ($M$ in the language of reference~\cite{RGF:2011}, $N$ in the language of this note), but there is little purpose in doing so --- the key insight is this: Once the data are assigned to ordinal boxes, hybrid geometric/power laws drop out automatically and straightforwardly by maximizing the Shannon entropy subject to the two very simple constraints $\langle \ln n\rangle=\chi$ and $\langle n\rangle=\mu$.

%------------------------------------------------------------------------------------------------------------------------------------------
\section{Zipf’s law: geometric version in infinite state space}\label{S:zipf2}
%------------------------------------------------------------------------------------------------------------------------------------------

If for observational or theoretical reasons one is \emph{certain} that $z=1$, (Zipf’s law), but for whatever reason feels a finite state space cutoff $N$ is inappropriate, then a geometric version of Zipf’s law can be extracted from the hybrid model. Setting $z=1$ the model reduces as follows: The state space is now $n\in\{1,2,3,\dots \}$ while
\begin{eqnarray}
\hat S &=& - \left( \sum_{n=1}^\infty p_n \ln n - \chi\right) + \ln w \left( \sum_{n=1}^\infty p_n n - \mu\right)  
\nonumber\\
&& \qquad - (\ln Z-1)  \left( \sum_{n=1}^\infty p_n  -1\right)
-\sum_{n=1}^\infty p_n \ln p_n.
\end{eqnarray}
This is completely equivalent to maximizing
\begin{eqnarray}
\fl
\hat S &=&  \chi  + \ln w \left( \sum_{n=1}^\infty p_n n - \mu\right)  
 - (\ln Z-1)  \left( \sum_{n=1}^\infty p_n  -1\right)
 -\sum_{n=1}^\infty p_n \ln (n \; p_n),
\end{eqnarray}
but writing the quantity to be maximized in this way hides the role of the Shannon entropy. 
Varying with respect to the $p_n$ yields
\begin{equation}
- \ln n + n \ln w - \ln Z - \ln p_n = 0,
\end{equation}
with solution
\begin{equation}
p_n = {1\over |\ln(1-w)|} \; {w^n\over n}; \qquad Z = |\ln(1-w)|; \qquad w \in (0,1).
\end{equation}
Note the normalizing function is now extremely simple --- the natural logarithm. 
Then
\begin{equation}
\chi(w) \equiv \langle \ln n \rangle= {1\over  |\ln(1-w)|}\; \sum_{n=1}^\infty { w^n \ln n\over n}; \qquad w \in(0,1),
\end{equation}
while
\begin{equation}
\mu(w) \equiv \langle n \rangle= {1\over  |\ln(1-w)|}\; \sum_{n=1}^\infty { w^n} =  {w\over (1-w) |\ln(1-w)|}.
\end{equation}
Furthermore, at maximum entropy %%%%%%%%%%%%%%
\begin{eqnarray}
\hat S (w)= S(w) &=& - \sum_{n=1}^\infty  p_n \ln p_n  
=  \ln  |\ln(1-w)| + \chi(w) -\mu(w) \ln w.
\end{eqnarray}
This is a 1-parameter model with a geometrical cutoff, which for $w\lesssim1$, (or more precisely $w\to 1^-$), approximates the naive un-normalizable Zipf law with arbitrary accuracy.

%
%------------------------------------------------------------------------------------------------------------------------------------------
\section{Very general Gibbs-like model}\label{S:gibbs}
%------------------------------------------------------------------------------------------------------------------------------------------
Let us now consider an arbitrary number of constraints of the form
\begin{equation}
\langle \ln g_i(n) \rangle \equiv \sum_n p_n \ln g_i(n) = \chi_i; 
\qquad i\in (1,\#_g), 
\end{equation}
and
\begin{equation}
\langle f_a(n) \rangle \equiv \sum_n p_n f_a(n) = \mu_a; 
\qquad a \in(1,\#_f).
\end{equation}
One could always transform a $g$-type constraint into an $f$-type constraint or vice versa, but as we shall soon see there are advantages to keeping the logarithm explicit.
Applying the maximum entropy principle amounts to considering
\begin{eqnarray}
\hat S &=& - \sum_i z_i \left( \sum_n p_n \ln g_i(n) - \chi\right) -\sum_a \beta_a  \left( \sum_n p_n f_a(n) - \mu\right)  
\nonumber\\
&& \qquad - (\ln Z-1)  \left( \sum_n p_n  -1\right) 
-\sum_n p_n \ln p_n,
\end{eqnarray}
where with malice aforethought we have now relabeled the Lagrange multipliers for the $f$ constraints as follows: $\ln w \to - \beta$. 
Then maximizing over the $p_n$ we have the extremality condition
\begin{equation}
- \sum_i z_i \ln g_i(n) - \sum_a \beta_a \; f_a(n) - \ln Z - \ln p_n = 0,
\end{equation}
with explicit solution
\begin{equation}
p_n = {1\over Z} \; \left\{ \prod_i g_i(n)^{-z_i}\right\} \; \exp\left\{-\sum_a \beta_a\; f_a(n) \right\};  
\end{equation}
where now the normalizing constant is
\begin{equation}
 Z(\vec z,\vec\beta) =  \sum_n \left[ \left\{\prod_i g_i(n)^{-z_i}\right\} \; \exp\left\{-\sum_a \beta_a\; f_a(n) \right\} \right].
\end{equation}
This can be viewed as a generalization/modification of the Gibbs distribution where we have explicitly pulled out some of the constraints (the $g$-type constraints) to make them look power-law-like, while the remaining constraints (the $f$-type constraints) are left in Boltzmann-like form. 
Then
\begin{equation}
\chi_i(\vec z,\vec\beta) \equiv \langle \ln g_i(n) \rangle= \sum_n p_n  \ln g_i(n) = - {\d \ln Z\over \d z_i},
\end{equation}
while
\begin{equation}
\mu_a(\vec z,\vec\beta) \equiv \langle f_a(n) \rangle= \sum_n p_n \, f_a(n) =   -{\d \ln Z\over \d \beta_a}.
\end{equation}
Furthermore, at maximum entropy
\begin{eqnarray}
\fl
\hat S(\vec z,\vec\beta)= S(\vec z,\vec\beta) &=& - \sum_n p_n \ln p_n =  
%\\
%&=&
 \ln Z(\vec z,\vec\beta) + \sum_i z_i \; \chi_i(\vec z,\vec\beta) + \sum_a \beta_a \; \mu_i(\vec z,\vec\beta).
\end{eqnarray}
It is only once one specifies particular choices for the functions $f_a$ and $g_i$ that the model becomes concrete, and only at that stage might one need to focus on particular special functions of mathematical physics. The model is extremely general --- the drawback is that, because it can fit almost anything, it can ``explain’’ almost anything, and so can predict almost nothing. 

%------------------------------------------------------------------------------------------------------------------------------------------
\section{Non-Shannon entropies}\label{S:non-shannon}
%------------------------------------------------------------------------------------------------------------------------------------------
Shannon’s entropy is by far the best motivated of the entropy functions infesting the literature. Without making any particular commitment to the \emph{advisability} of doing so, we can certainly ask what happens if we apply maximum entropy ideas to non-Shannon entropies (such as the R\'enyi~\cite{Renyi} or Tsallis~\cite{Tsallis} entropies and their generalizations). Let us define an entropic zeta function by
\begin{equation}
\zeta_S (s)= \sum_n (p_n)^s,
\end{equation}
which certainly converges for $s\geq 1$ and \emph{may} converge on a wider region.
Then
\begin{equation}
S_\mathrm{Renyi}(1+a) = -{\ln \zeta_S(1+a)\over a}; \qquad S_\mathrm{Tsallis}(1+a) = {1-\zeta_S(1+a)\over a},
\end{equation}
where in both cases the Shannon entropy is recovered in the limit $a\to0$.
More generally let us consider a generalized entropy of the form
\begin{equation}
S(a) = -f\left( \zeta_S(1+a)\right),
\end{equation}
for an arbitrary smooth function $f(\cdot)$. Let us further impose a constraint on the $b^\mathrm{th}$ moment
\begin{equation}
\langle n^b \rangle \equiv \sum_n p_n \; n^b = \mu_b.
\end{equation}
With power laws being explicitly built in as input into both the generalized entropy and the constraint, it is perhaps not too surprising that we will manage to get power laws dropping out. 
One is now interested in maximizing
\begin{eqnarray}
\hat S &=& \lambda \left(\sum_n p_n \; n^b - \mu_b\right) + w  \left( \sum_n p_n  -1\right)
- f\left( \zeta_S(1+a)\right).
\end{eqnarray}
Varying the $p_n$ leads to
\begin{equation}
\lambda n^b + w-  f'\left( \zeta_S(1+a)\right) \; (1+a) (p_n)^a =0,
\end{equation}
with solution
\begin{equation}
p_n = {(w+\lambda \; n^b)^{1/a}\over Z}; \qquad Z = \sum_n (w+\lambda \; n^b)^{1/a}.
\end{equation}
(An overall factor of $(1+a)\; f'\left( \zeta_S(1+a)\right)$ simply drops out of the calculation.)
If the number of states is finite, then we cannot \emph{a priori} discard the parameter $w$, and we have derived a distorted power law. (The probability distribution then interpolates between a pure power law for $w=0$ and a uniform distribution for $w=\infty$.) 
If on the other hand, the number of states is infinite then normalizability enforces $w\to0$, and $\lambda$ factors out. We then have a pure power law
\begin{equation}
p_n \to {n^{b/a}\over Z}; \qquad Z \to \sum_n  n^{b/a}.
\end{equation}
If the state space is the positive integers $n\in\{1,2,3,\dots\}$ then $Z\to\zeta(-b/a)$, the Riemann zeta function, and the sum converges only for $-b/a > 1$. So one of the two parameters ($a$, $b$) must be negative. 
In this situation
\begin{equation}
\mu_b = \langle n^b \rangle = {\zeta(-b-b/a)\over\zeta(-b/a)},
\end{equation}
now requiring both $-b-b/a>1$ and $-b/a>1$, 
while at maximum entropy
\begin{equation}
\hat S = S \to  f\left( \zeta_S(1+a)\right) =  f\left( {\zeta(-b-b/a)\over \zeta(-b/a)^{1+a}} \right).
\end{equation}
So yes, one can also extract power laws from maximum entropy applied to non-Shannon entropies, (in particular, the generalized R\'enyi--Tsallis entropies), but the process is (at best) rather clumsy, and seems an exercise in overkill. Apart from the whole question of whether or not non-Shannon entropies are particularly interesting, one should ask whether the result is particularly useful? This derivation does not seem to be in any way an improvement over the simpler one based directly on the Shannon entropy, so its utility is dubious.

%------------------------------------------------------------------------------------------------------------------------------------------
\section{Summary and Discussion}\label{S:discussion}
%------------------------------------------------------------------------------------------------------------------------------------------

The main point of this article is that power laws, (and their variants, including hybrid geometric/power laws), have a very natural and straightforward interpretation in terms of the maximum entropy formalism pioneered by Jaynes. The key to obtaining a pure power law in the simplest possible manner lies in maximizing the Shannon entropy while imposing the simple constraint $\langle \ln n\rangle=\chi$.  Depending on other features of the specific model under consideration, detailled analysis leads to certain of the special functions of mathematics, (the Riemann zeta function, generalized harmonic functions, poly-logarithms, or even ordinary logarithms), but these are relatively well-known mathematical objects, which are still tolerably simple to deal with. 

Adding additional features (finite size state space, extra constraints) can (and typically will) modify both the functional form and the normalization constants appearing in the probability distribution. 
As always there is a trade-off between simplicity and flexibility. A more complicated model (with more free parameters) has a more flexible probability distribution, but this comes at a real (if often unacknowledged) cost in terms of internal complexity. A rather general Gibbs-like model is laid out and briefly discussed.  We also briefly discuss applying maximum entropy ideas to non-Shannon entropies (such as the R\'enyi and Tsallis entropies). There is very definitely a trade-off in both elegance and plausibility, and I would argue strongly that the simplest and most elegant model consists of the Shannon entropy, a constraint on $\langle \ln n\rangle=\chi$, and a trivial normalization constraint on the sum of probabilities. 

%\clearpage
The fact that the logarithmic average $\langle \ln n \rangle$ plays such an important role in power laws seems to have a connection with the fact that logarithmic scales are ubiquitous in classifying various natural and social phenomena. For instance:
\begin{itemize}
\item Stellar magnitudes are logarithmic in stellar luminosity.
\item Earthquake magnitudes (modified Richter scale) are logarithmic in energy release.
\item Sound intensity decibels are logarithmic in pressure.
\item The acidity/alkalinity pH scale is logarithmic in hydrogen ion concentration.
\item Musical octaves are logarithmic in frequency.
\item War severity can be characterized as being logarithmic in casualty count~\cite{quarrels}.
\end{itemize}
In many cases the utility of a logarithmic scale can be traced back to an approximate logarithmic sensitivity in human perceptual systems,  but it is very easy to confound cause and effect. After all, in the presence of power-law distributed external stimuli, there is a significant disadvantage in having the human perceptual system overwhelmed by large numbers of low-impact events, suggesting an evolutionary pressure towards suppressing sensitivity to low-impact events.  \emph{This suggests that logarithmic sensitivity in human (and animal) perceptual systems is evolutionarily preferred for those senses that are subject to an external bath of power-law distributed stimuli}. 

Fortunately, for the purposes of applying maximum entropy ideas one does not need to know which is the cause and which is the effect --- one is ``merely’’ using Bayesian principle to \emph{estimate} underlying probabilities in the presence of limited knowledge; for current purposes this is most typically the single piece of information that $\langle \ln n\rangle=\chi$.

%--------------------------------------------------------------------------------------------------------------------------
\ack
%--------------------------------------------------------------------------------------------------------------------------

MV acknowledges support via the Marsden Fund, and via a James Cook Fellowship, both administered by the Royal Society of New Zealand.

%------------------------------------------------------------------------------------------------------------------------------------------
\section*{References}
%------------------------------------------------------------------------------------------------------------------------------------------

%------------------------------------------------------------------------------------------------------------------------------------------

\begin{thebibliography}{69}
%------------------------------------------------------------------------------------------------------------------------------------------

\bibitem{Zipf:1935}
George K Zipf, \emph{The Psychobiology of Language}, (Houghton-Mifflin, Cambridge, USA, 1935)

\bibitem{Zipf:1949}
George K Zipf, \emph{Human Behavior and the Principle of Least Effort}, (Addison-Wesley, 1949).

\bibitem{SMS:2009}
Alexander Saichev, Yannick Malevergne, and Didier Sornette, \emph{Theory of Zipf's law and beyond}, Lecture Notes in Economics and Mathematical Systems, Volume 632, Springer (November 2009), ISBN 978-3-642-02945-5.

\bibitem{Newman:2005}
M E J  Newman, ``Power laws, Pareto distributions and Zipf's law”. \\
Contemporary Physics {\bf46} (2005) 323--351 [arXiv:cond-mat/0412004]. \\
doi:10.1080/00107510500052444.

\bibitem{Newman:2009}
Aaron Clauset, Cosma Rohilla Shalizi, and M E J Newman, \\
``Power-law distributions in empirical data”, \\
SIAM Review {\bf51} (2009) 661--703 [arXiv:0706.1062v2]. doi:10.1137/070710111.

\bibitem{Java}
Gareth Baxter, Marcus Frean, James Noble, Mark Rickerby, Hayden Smith, Matt Visser, Hayden Melton and Ewan Tempero,
“Understanding the Shape of Java Software”, 
OOPSLA 2006, 
Proceedings of the 21st annual ACM SIGPLAN conference on Object-oriented programming systems, languages, and applications. Pages 397-412.\\ 
Editors: Peri L. Tarr and  William R. Cook. (ACM Press, New York).\\
doi: 10.1145/1167515.1167507;  ISBN:1-59593-348-4

\bibitem{quarrels}
Lewis Fry Richardson, \emph{Statistics of deadly quarrels}, \\
(Boxwood Press, Pacific Grove, CA, 1960). ISBN	0910286108; ISBN13 9780910286107.

\bibitem{RGF:2011} Seung Ki Baek, Sebastian Bernhardsson, Petter Minnhagen, ``Zipf's law unzipped’’, \\
New J. Phys. {\bf4} (2011) 043004 [arXiv:1104.1789 [physics.soc-ph]]. \\
doi:10.1088/1367-2630/13/4/043004.

\bibitem{Jaynes:1957a}
E T Jaynes, ``Information Theory and Statistical Mechanics’’, \\
Physical Review Series II {\bf106} (1957) 620--630. doi:10.1103/PhysRev.106.620.
 
 \bibitem{Jaynes:1957b}
E T Jaynes, ``Information Theory and Statistical Mechanics II’’, \\
Physical Review Series II {\bf108} (1957) 171--190. doi:10.1103/PhysRev.108.171.

\bibitem{Jaynes:1968}
E T Jaynes, ``Prior Probabilities”, \\
IEEE Transactions on Systems Science and Cybernetics {\bf4} (1968) 227--241. \\
doi:10.1109/TSSC.1968.300117.

\bibitem{Jaynes:1988}
E T Jaynes, ``The Relation of Bayesian and Maximum Entropy Methods’’, \\
in \emph{Maximum-Entropy and Bayesian Methods in Science and Engineering}, Vol. 1, \\
Editors: G. Erickson, C.R. Smith (Kluwer Academic Publishers, 1988), p. 25--29.\\
ISBN-10: 9027727937 ISBN-13: 978-9027727930


\bibitem{Jaynes:2003}
E T Jaynes, \emph{Probability Theory: The Logic of Science}, (Cambridge University Press, 2003). \\ pp.~351--355. ISBN 978-0521592710

\bibitem{Cristelli:2012}
M. Cristelli, M. Batty, and L. Pietronero, “There is More than a Power Law in Zipf”,  \\
Scientific Reports {\bf2} (2012) 812.
doi: 10.1038/srep00812

\bibitem{Hernando:2012a}
Hernando A, Plastino A, and Plastino AR, “MaxEnt and dynamical information”, \\
Eur Phys J B {\bf85} (2012) 147.
[arXiv:1201.0889 [physics.data-an]]. \\
doi: 10.1140/epjb/e2012-30009-3

\bibitem{Hernando:2012b}
Hernando A, Plastino A, “Variational principle underlying scale invariant social systems”, \\
Eur Phys J B {\bf85} (2012) 293.
[arXiv:1204.2420 [stat.AP]]. doi: 10.1140/epjb/e2012-30313-x

\bibitem{Hernando:2012c}
Hernando A, Plastino A, “The thermodynamics of urban population flows”, \\
Phys Rev E {\bf86} (2012) 066105.
[arXiv:1206.7020 [physics.soc-ph]]. \\
doi: 10.1103/PhysRevE.86.066105

\bibitem{Hernando:2012d}
Hernando A, Hernando R, Plastino A, and Plastino AR, \\
“The workings of the maximum entropy principle in collective human behaviour”, \\
J R Soc Interface (2013) 10:20120758.
[arXiv:1201.0905 [stat.AP]]. doi: 10.1098/rsif.2012.0758

\bibitem{barcelona}
R. Pastor--Satorras and J. Wagensberg,\\
“The maximum entropy principle and the nature of fractals”,
Physica A 251 (1998) 291–302

\bibitem{Shannon:1948}
Claude E Shannon, ``A Mathematical Theory of Communication’’, \\
Bell System Technical Journal {\bf27} (3) (July/October 1948)  379--423. 

\bibitem{Shannon:1949}
Claude E Shannon, Warren Weaver, ``The Mathematical Theory of Communication’’, \\
University of Illinois Press, 1949. ISBN 0-252-72548-4

\bibitem{Renyi}
Alfréd Rényi,  “On measures of information and entropy”, \\
Proceedings of the 4th Berkeley Symposium on Mathematics, Statistics and Probability 1960. (Published 1961). pp. 547–561.

\bibitem{Tsallis}
Constantino Tsallis, “Possible generalization of Boltzmann-Gibbs statistics”,\\
 Journal of Statistical Physics 52 (1988) 479–487. doi:10.1007/BF01016429.


\bibitem{pure}
G H Hardy, \emph{A course in pure mathematics},\\
 (Cambridge University Press, England, 1908, reprinted 2006).   

\bibitem{dirichlet-series}
G H Hardy and  Marcel Riesz,   \emph{The general theory of Dirichlet’s series}, \\
(Cambridge University Press, England, 1915). 
Republished by Cornell University Library 1991.

\bibitem{shilov}
Georgi E Shilov, \emph{Elementary real and complex analysis}, (Dover, New York, 1996). 

\bibitem{Havil} 
Julian Havil, \emph{Gamma: Exploring Euler’s constant},  (Princeton University Press, Princeton, 2003).

%------------------------------------------------------------------------------------------------------------------------------------------
\end{thebibliography}
\end{document}